\documentclass[%
 reprint,
superscriptaddress,
 amsmath,amssymb,
 aps,
]{revtex4-2}

\usepackage{graphicx}
\usepackage{dcolumn}
\usepackage{bm}
\usepackage{xcolor}
\usepackage{float}

\begin{document}

\title{First-principles study of dielectric properties of ferroelectric perovskite oxides with on-site and inter-site Hubbard interactions}

\author{Min Chul Choi}
\affiliation{Department of Physics and Integrative Institute of Basic Sciences, Soongsil University, Seoul, 06978, Korea}

\author{Wooil Yang}
\email{Email: yspacefirst@kias.re.kr}
\author{Young-Woo Son}
\email{Email: hand@kias.re.kr}
\affiliation{Korea Institute for Advanced Study, Seoul 02455, Korea}
\author{Se Young Park}%
\email{Email: sp2829@ssu.ac.kr}
\affiliation{Department of Physics and Integrative Institute of Basic Sciences, Soongsil University, Seoul, 06978, Korea}
\affiliation{Origin of Matter and Evolution of Galaxies (OMEG) Institute, Soongsil University, Seoul, 06978, Korea}

\date{\today}

\begin{abstract}
We study the atomic and electronic structures of ferroelectric perovskite oxides, BaTiO$_3$, LiNbO$_3$, and PbTiO$_3$ using {\it ab initio} extended Hubbard functionals in which the on-site and inter-site Hubbard interactions are determined self-consistently, adapted from the pseudohybrid density functional proposed by Agapito–Curtarolo–Buongiorno Nardelli. Band structures, ferroelectric distortions, polarization, Born effective charges, and switching barriers are calculated with extended Hubbard functionals, that are compared with those using local density approximation (LDA), generalized gradient approximation (GGA), and Hybrid (HSE06) functionals. The properties of all three compounds calculated by extended Hubbard functionals are in good agreement with experimental data. We find a substantial increase in band gaps due to the inter-site Coulomb interactions, which show better agreement with $GW$ results compared to those from LDA and GGA functionals. The crucial role of the inter-site Coulomb interactions in restoring the suppressed polar instability, which is computed when only the on-site Hubbard interactions are considered, is also highlighted. 
Overall, we find that the properties calculated using our extended Hubbard functionals exhibit trends similar to those obtained with the HSE06 functional, while reducing computational costs by over an order of magnitude. Thus, we propose that the current method is well-suited for high-throughput calculations for perovskite oxides, offering significantly improved accuracy in computing band gap and other related physical properties such as the shift current photovoltaic effect and band alignments in ferroelectric heterostructures.
\end{abstract}

\maketitle

\section{\label{sec:intro}Introduction}
Materials showing switchable polarization have been extensively investigated due to various applications such as high-information density memory, high-K dielectrics, actuators, and transducers for energy harvesting \cite{scott1989,ahn2004,dawber2005,rabe2007,scott2007}. New directions in utilizing switchable polarization to control the material properties are also actively investigated. For example, the shift current photovoltaic effect \cite{Von1981,Sipe2000,Tan2016} and non-linear Hall effect \cite{Sodemann2015,Ma2019,Lee2024} require the inversion symmetry breaking, in which the direction of non-linear current can be controlled by changing the polarization direction \cite{nakamura2017,kang2023}. Moreover, various heterojunctions utilizing interfacial interactions with ferroelectric materials exhibit spontaneous symmetry broken phases \cite{Yadav2016,Das2019,Wang2018}, multiferroelectricity \cite{Mundy2016}, and controllable magnetic properties by electric fields \cite{Sahoo2007,Shirahata2015,Eom2023}.  

The growing attention to utilizing ferroelectricity in a wide range of physical phenomena and related applications requires accurate prediction of the properties of materials with switchable polarization. In particular, finding computationally efficient ways to evaluate the properties of a large number of materials is increasingly important. This need is amplified by growing relevance of high-throughput calculations combined with machine-learning-based materials predictions~\cite{Behler2016,Bedolla2020}, which are widely used to identify materials of desired properties. First-principles density functional theory (DFT) has been a main machinery in the high-throughput prediction of ferroelectric materials \cite{Schmidt2019,Ricci2024}, in which properties including the polar distortions and the switching polarization can be accurately calculated. However, compared with these properties, the predicted band gap values are significantly smaller than the experimental values for the majority of exchange-correlation functionals that are used in high-throughput calculations. Accurate prediction of band gap values is crucial in determining physical properties, such as frequency of absorbed light in shift-current photovoltaic effects and band alignments and associated charge transfer of ferroelectric heterojunction.  Therefore, an accurate and efficient scheme that captures both ferroelectric properties and band gap predictions by proper treatment of Coulomb interaction would be useful to the acquisition of large, reliable materials data sets, which in turn can facilitate machine learning-based studies.

Traditionally exchange-correlation functionals with the local density approximation (LDA)~\cite{Kohn1965} and the generalized gradient approximation (GGA)~\cite{Perdew1996} are widely used to predict ferroelectric properties. With the development of the meta-GGA and hybrid functionals, comparisons of the ferroelectric properties depending on the choice of exchange-correlation functionals are documented \cite{Bilc2008,Zhang2017,Paul2017,Wahl2008,Sun2016}. In particular,  the strongly constrained and appropriately normed (SCAN) meta-GGA functional \cite{Sun2015} provides an accurate prediction of ferroelectric properties of conventional ferroelectrics \cite{Paul2017} but with underestimated band gap values \cite{Bilc2008,Zhang2017,Wahl2008}. This is due to self-interaction error (SIE) \cite{Perdew1981,Mori2006}, which has been a major challenge in improving the band gap predictions~\cite{Schilfgaarde2006}. The hybrid functional (HSE06)~\cite{heyd2006}, in which the self-interaction correction of screened Hartree-Fock interaction is mixed with the exchange part of GGA exchange functional, is shown to successfully increase the accuracy of calculated band gap values with well reproduced ferroelectric properties \cite{Bilc2008,Zhang2017}. The band-gap values can also be corrected by the $GW$ method by evaluating the screened Coulomb interaction within random phase approximation (RPA)~\cite{Hybertsen1985,Schilfgaarde2006,Shishkin2007prl,jiang2010prb,Lany2013prb,Hinuma2014prb,Jiang2016prb}. However, both hybrid functional and $GW$ methods require calculation times that are two to three orders of magnitude longer than those of LDA and SCAN, which makes these calculation schemes challenging for large-scale calculations.   
 
An alternative approach to treat SIE is considering the Coulomb interaction among a small set of correlated orbitals, motivated by the Hubbard model, such as DFT+$U$ methods, which account for the on-site Coulomb interaction within $d$ or $f$ orbitals \cite{Anisimov1991,Liechtenstein1995,Dudarev1998}. The extension of the methodology including the inter-site Coulomb interaction based on the extended Hubbard model, namely the DFT+$U$+$V$ method, is proposed~\cite{Campo2010}, offering reliable predictions of semiconductor band gaps with computational efficiency comparable to conventional DFT~\cite{lee2020,Tancogne2020,Timrov2020prr,Mahajan2021prm,Timrov2022prxe}. Applying these schemes to high-throughput calculations presents challenges in evaluating the interaction parameters. Determining on-site and inter-site Hubbard parameters involves methods such as constrained DFT~\cite{Dederichs1984, Hybertsen1989prb, Pickett1998prb, Solovyev2005, Nakamura2006, Shishkin2016prb, Timrov2018, Timrov2021} and constrained RPA calculations~\cite{Springer1998, Kotani2000jpcm, Aryasetiawan2004, Aryasetiawan2006}. These methods require additional computations, such as careful control of orbital occupancy, Wannierization, and linear-response calculations for correlated orbitals. As a result, automating these processes for large-scale materials searches remains a complex challenge.

Recently, the pseudohybrid Hubbard density functional proposed by Agapito-Curtarolo-Buongiorno Nardelli (ACBN0)~\cite{agapito2015} enables a direct determination of $U$ within self-consistent field calculations. As the generalization of the ACBN0 functional considering both on-site and inter-site Hubbard parameters \cite{lee2020,Tancogne2020}, the extended ACBN0 functional self-consistently determines the Hubbard parameters $U$ and $V$. It is shown that the extended ACBN0 scheme accurately predicts band gap values for a wide range of semiconductors and insulators with calculation costs comparable with local or semilocal exchange-correlation functionals~\cite{lee2020,Tancogne2020}. 

Given that the extended ACBN0 scheme provides an effective way to correct the underestimated band gap values, confirming the validity of the methodology in predicting the ferroelectric properties is an important question for efficient high-throughput-calculation-based searches of ferroelectric materials. To the best of our knowledge, there has been no systematic investigation of the properties of ferroelectric materials using the ACBN0 scheme. In addition, the effect of including the inter-site Coulomb interaction (difference between ACBN0 and extended ACBN0 scheme) on the ferroelectric properties has not yet been studied, which is also an important question, as the inter-site Coulomb interaction is required to stabilize the polar phase in BaTiO$_3$, recently reported in a constrained DFT-based DFT+$U$+$V$ study~\cite{Gebreyesus2023}.    

In this paper, we present a systematic investigation of the structural, electronic, and ferroelectric properties of ferroelectric oxides using LDA, GGA, HSE06, and both ACBN0 and its extension. Here, we select three representative conventional perovskite ferroelectrics, BaTiO$_3$ (BTO), PbTiO$_3$ (PTO), and LiNbO$_3$ (LNO), for which ferroelectric properties have been extensively studied and compared for various exchange-correlation functionals \cite{Bilc2008,Zhang2017,Paul2017,Wahl2008,Sun2016}. In particular, we focus on the role of the inter-site Coulomb interaction in polarization switching and band gap predictions. Our study shows that  properties calculated with the DFT+$U$+$V$ method, based on the extended ACBN0 functional, are in good agreement with experimental results, offering much improved accuracy in band gap predictions. Overall, the calculated values are close to those obtained from the hybrid functionals, yet the calculation time is drastically reduced, making it comparable to local and semilocal exchange functionals. We hope that the extended ACBN0 method can be employed to construct extensive and accurate databases by high-throughput calculations, thus contributing to the exploration and research of database-driven studies.

\section{\label{sec:methods}Methods}
We provide a brief overview of the DFT method incorporating both $U$ and $V$ interactions~\cite{Dudarev1998,Campo2010}, and discuss a recently developed pseudohybrid functional for inter-site interactions~\cite{lee2020,Tancogne2020}. The total energy with the extended Hubbard energy functionals can be written as 
\begin{equation}
E_\textrm{tot}=E_\textrm{DFT}+E_{U+V}-E_\textrm{dc},
\label{Eq:tot}
\end{equation}
where $E_\textrm{DFT}$ represents the standard DFT energy obtained with any local or semilocal density functional like LDA or GGA, $E_{U+V}$ the on-site and inter-site electron interaction energy, and $E_\textrm{dc}$ the double counting of electron interaction energy which already exists in $E_\textrm{DFT}$. 

The expression for $E_{U+V}$ within the mean field approximations~\cite{Campo2010,lee2020} is
\begin{eqnarray}
E_{U+V}&=&\sum_{I}
\left[
{
\frac{U^I}{2}}
\sum_{ii',\sigma}
{n^{I\sigma}_{i}n^{I\bar{\sigma}}_{i'}} 
+{\frac{U^I-J^I}{2}}\sum_{i \neq i',\sigma}
{n^{I\sigma}_{i}n^{I\sigma}_{i'}} 
\right]
\nonumber \\
&+&\sum_{\{I,J\}}
\sum_{ij,\sigma\sigma'}
{\frac{V^{IJ}}{2}}
\left[
{n^{I\sigma}_{i}n^{J\sigma'}_{j}}-\delta_{\sigma\sigma'}{n^{IJ\sigma}_{ij}n^{JI\sigma'}_{ji}}
\right],
\label{eq:el-el}
\end{eqnarray}
where the $\{I,J\}$ indicates the pairs of atoms $I$ and $J$ within a cutoff distance and
$\bar{\sigma}$ is an opposite spin orientation 
to $\sigma$. 
In Eq.~\eqref{eq:el-el}, the generalized occupation number is written as 
\begin{eqnarray}
n^{IJ\sigma}_{ij}&=& n^{I,n_I,l_I,J,n_J,l_J,\sigma}_{ij}\nonumber \\
&\equiv& \sum_{m{\bf k}}f_{m\bf k}^\sigma
\langle \psi_{m{\bf k}}^\sigma |\phi^{I,n_I,l_I}_i \rangle \langle \phi^{J,n_J,l_J}_j |\psi_{m{\bf k}}^\sigma\rangle \nonumber\\
&=& \sum_{m{\bf k}}f_{m\bf k}^\sigma
\langle \psi_{m{\bf k}}^\sigma |\phi^{I}_i \rangle \langle \phi^{J}_j |\psi_{m{\bf k}}^\sigma\rangle,
\label{Eq:GO}
\end{eqnarray}
where $f_{m\bf k}^\sigma$ is the Fermi-Dirac function of 
the Bloch state $|\psi_{m\bf k}^\sigma\rangle$ 
of the $m$-th band at a momentum ${\bf k}$ and the principle ($n_{I(J)}$), azimuthal ($l_{I(J)}$), angular ($i(j)$), and spin ($\sigma$) quantum numbers  
of the $I(J)$-th atom are implicitly written in the last line of Eq.~\eqref{Eq:GO}. $|\phi^I_i\rangle$ represents the projector for atomic orbitals and the L\"{o}wdin orthonormalization is used to rule out the latent double counting of Hubbard corrections in overlap regions between atoms~\cite{lee2020}. We note that when $I=J$ and $i=j$, Eq.~\eqref{Eq:GO} becomes a usual density matrix for the single orbital Hubbard model, $n^{I\sigma}_i\equiv n^{II\sigma}_{ii}$.

The expression for $E_\textrm{dc}$~\cite{Campo2010,lee2020,Tancogne2020} is
\begin{eqnarray}
E_\textrm{dc} &=& \sum_{I}
\left[
{\frac{U^I}{2}}N^{I}(N^{I}-1)-\sum_\sigma\frac{J^{I}}{2}N^{I\sigma}(N^{I\sigma}-1 )
\right]
\nonumber \\ 
& &+\sum_{\{I,J\}}{\frac{V^{IJ}}{2}}N^{I}N^{J},
\label{eq:dc}
\end{eqnarray}
where $N^{I\sigma}=\sum_{i}n^{I\sigma}_{i}$ and $N^{I}=\sum_{\sigma}N^{I\sigma}$. The double counting term in Eq.~\eqref{eq:dc} was adopted from fully localized limit (FLL) expression~\cite{Dudarev1998}. This form represents the mean-field approximation of $E_\textrm{Hub}=E_\textrm{$U$+$V$}-E_\textrm{dc}$ in the limit where the occupation number of atomic orbitals is either fully occupied or empty.

Then, $E_{U+V}$ can be expressed in an invariant form under rotation of the orbitals of the same atomic site:

\begin{equation}
E_{U+V} =\sum_{I, \sigma} \frac{U^I_\text{eff}}{2}\text{tr}
\left[
{\bf n}_{I\sigma}-{\bf n}^2_{I\sigma}
\right]
-\sum_{\{I,J\}, \sigma}\frac{V^{IJ}}{2}\text{tr}
\left[
{\bf n}_{IJ\sigma} {\bf n}_{JI\sigma}
\right]
\label{Eq:Hub}
\end{equation}
where $U^I_\text{eff}=U^I-J^I$, 
trace is over orbitals, and
${\bf n}_{I\sigma}$ and ${\bf n}_{IJ\sigma}$ are matrix notations for $n^{I\sigma}_i$ and $n^{IJ\sigma}_{ij}$, respectively. This is a generalization of Dudarev formulation of DFT+$U$~\cite{Dudarev1998} to DFT+$U$+$V$ proposed by Campo Jr and Cococcioni~\cite{Campo2010}.
We determine the Hubbard parameters of $U^I$, $J^I$ and $V^{IJ}$ self-consistently based on recent developments~\cite{lee2020,Tancogne2020} that extend the method of ACBN0~\cite{agapito2015} exploiting partial projections of unresticted Hartree-Fock energy~\cite{Mosey2007,Mosey2008}. 
Here we only consider the collinear spin configuration~\cite{yang2024}.
In these approaches, all Hubbard parameters can be expressed as rescaled bare Coulomb interactions of
$
(Ii,Jj|Kk,Ll)
\equiv\int d{\bf r}_1 d{\bf r}_2 
\frac{\phi^{I*}_i ({\bf r}_1)\phi^J_j({\bf r}_1)\phi^{K*}_k ({\bf r}_2)\phi^L_l ({\bf r}_2)}{|{\bf r}_1-{\bf r}_2|}
\label{Eq:ERI}
$
~\cite{lee2020} such that, 
\begin{eqnarray}
U_{I}&=&
 \sum_{ijkl,\sigma\sigma'} 
\frac{{ P}^{I\sigma}_{ij}{ P}^{I\sigma'}_{kl}}{N_{U_I}}
(Ii,Ij|Ik,Il),
\label{eq:uf}\\
J_I&=& \sum_{ijkl,\sigma\sigma'} 
\frac{
{P}^{I\sigma}_{ij}{P}^{I\sigma}_{kl}}
{{N}_{J_I}}
(Ii,Il|Ik,Ij) ,
\label{eq:jf}\\
V_{IJ}&=&
\sum_{ijkl,\sigma\sigma'}
\frac{{P}^{I\sigma}_{ij}{P}^{J\sigma'}_{kl}-\delta_{\sigma\sigma'}{P}^{IJ\sigma}_{il}{ P}^{JI\sigma}_{kj}}{{N}_{V_{IJ}}} 
\nonumber\\
& &\times(Ii,Ij|Jk,Jl),
\label{eq:vf}
\end{eqnarray}
where 
\begin{eqnarray}
{N}_{U_I} &=& \sum_{i\neq j,\sigma}n^I_{i\sigma}n^I_{j\sigma}+\sum_{ij,\sigma}
n^I_{i\sigma}n^I_{j\bar{\sigma}},\nonumber\\
{N}_{J_I} &=& \sum_{i\neq j,\sigma} 
n^I_{i\sigma}n^I_{j\sigma}, \nonumber \\
{N}_{V_{IJ}} &=&\sum_{ij,\sigma\sigma'}
\left[{n^{I\sigma}_{i}n^{J\sigma'}_{j}}-\delta_{\sigma\sigma'}
{n^{IJ,\sigma\sigma'}_{ij}n^{JI,\sigma'\sigma}_{ji}}
\right],
\end{eqnarray}
and 
\begin{equation}
{P}^{IJ\sigma}_{ij} = \sum_{m\mathbf{k}}f^\sigma_{m\mathbf{k}} {N}^{IJ\sigma}_{\psi_{m\mathbf{k}}} \langle \psi^{\sigma}_{m\mathbf{k}} | \phi^{I}_{i} \rangle \langle \phi^{J}_{j} | \psi^{\sigma}_{m\mathbf{k}} \rangle.
\label{eq:rdm}
\end{equation}
For renormalized density matrix~\cite{agapito2015} in Eq.~\eqref{eq:rdm}, we define renormalized occupation number for different atoms $I$ and $J$ that can be written as~\cite{lee2020},
\begin{eqnarray}{N}^{IJ\sigma}_{\psi_{m\mathbf{k}}} 
&=& \sum_{\{I \}}\sum_{i} \langle \psi^{\sigma}_{m\mathbf{k}} |  \phi^{I}_{i} \rangle \langle  \phi^{I}_{i} | \psi^{\sigma}_{m\mathbf{k}} \rangle \nonumber \\
& &+ \sum_{\{J \}}\sum_{j} \langle \psi^{\sigma}_{m\mathbf{k}} |  \phi^{J}_{j} \rangle \langle  \phi^{J}_{j} | \psi^{\sigma}_{m\mathbf{k}} \rangle, 
\label{eq:rom}
\end{eqnarray}
where $\{I(J)\}$ in summation represents all the orbitals with quantum number $n_{I(J)}$ and $l_{I(J)}$ of an atom type $I(J)$. For a pair of the same atoms or for the $U^I$ calculations, 
${P}^{I\sigma}_{ij}\equiv {P}^{II\sigma}_{ij}$ and 
Eq. \eqref{eq:rom} reduces to ${N}^{I\sigma}_{\psi_{m\mathbf{k}}} = \sum_{\{I \}}\sum_{i} \langle \psi^{\sigma}_{m\mathbf{k}} |  \phi^{I}_{i} \rangle \langle  \phi^{I}_{i} | \psi^{\sigma}_{m\mathbf{k}} \rangle$ which is nothing but a L\"{o}wdin charge of atom $I$ for a Bloch state of $\vert \psi^{\sigma}_{m\mathbf{k}} \rangle$.

\section{\label{sec:comp}Computational Details}
Structural, electronic, and ferroelectric properties are calculated using \textsc{Quantum Espresso (QE)}~\cite{Giannozzi2009,Giannozzi2017} software with norm-conserving (NC) pseudopotential. LDA \cite{Perdew1981}, GGA by Perdew-Burke-Ernzerhof revised for solids (PBEsol)~\cite{Perdew2008} are used. The self-consistent extended Hubbard parameters based on the ACBN0 method and its extension~\cite{agapito2015,lee2020} are obtained by the modified in-house QE package using L\"{o}wdin orthogonalized atomic orbitals as projectors. An energy cutoff for the NC pseudopotentials is 110 Ry (1496 eV). The Brillouin zone sampling is on a $\Gamma$-centered $k$-point grid of 8 $\times$ 8 $\times$ 8 for BTO and PTO, and 6 $\times$ 6 $\times$ 6 for LNO. The atomic positions are fully relaxed with a force threshold of 5 $\mu$Ry/$a_0$  (0.001 eV/\AA), with $a_0$ being the Bohr radius. For the HSE06 calculations, \textsc{Vienna Ab-initio Simulation Package (VASP)}~\cite{Kresse1995,Kresse1996} with the projector-augmented wave (PAW)~\cite{Blochl1994} is used with an energy cut-off of 500 eV. The $k$-point grid and force thresholds are set to the same values as the QE calculations. The polarization and the Born effective charges (BECs) are calculated using the Berry phase method~\cite{King1993} and the finite field approach~\cite{Nunes2001prb,Souza2002prl}, respectively. For the visualization of atomic structures, the VESTA software \cite{Momma2011} is used.

\section{\label{sec:results}Results}

\subsection{\label{subsec:struc}Atomic structures of ferroelectric BTO, PTO, and LNO}

\begin{figure}[htbp]
\includegraphics[width=0.6\columnwidth, angle=-0]{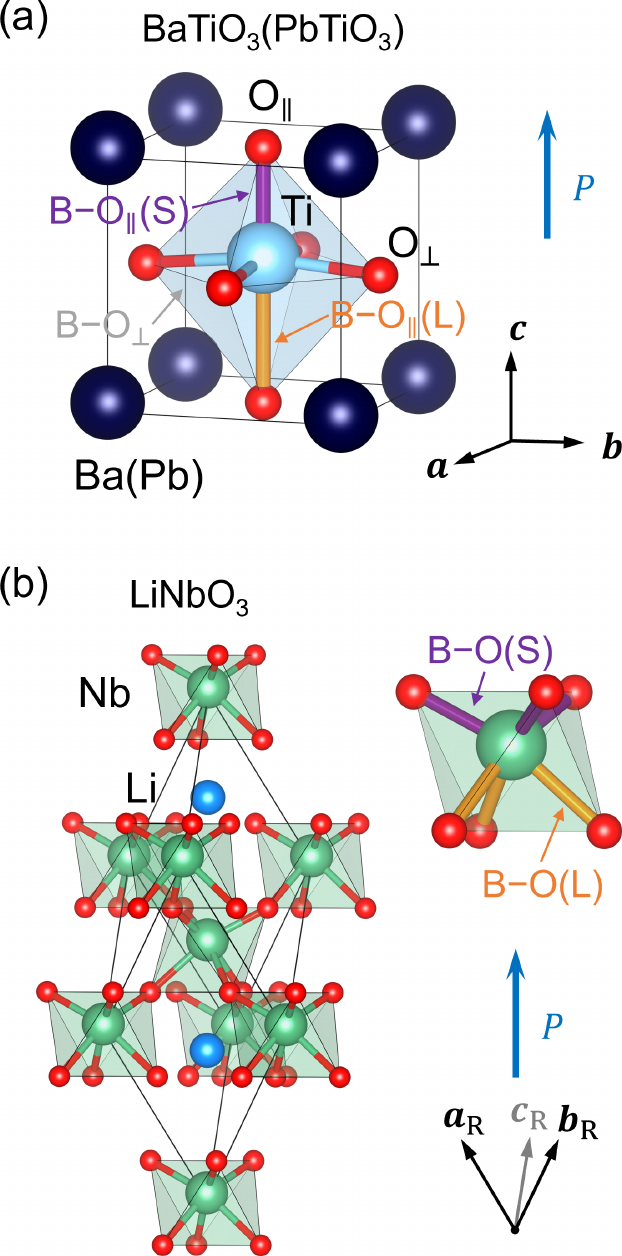}
\caption{\label{fig:atomic_str} Atomic configurations of BTO, PTO, and LNO in ferroelectric phase.  (a) BTO (PTO) atomic structures in tetragonal $P4mm$ space group with polarization along the $c$-axis (blue arrow). There are three distinct bondings between Ti-$d$ and O-$p$ orbitals due to the polar distortion. The inequivalent bondings between Ti and O atoms are denoted with different colors.  
(b) LNO atomic structures in rhombohedral $R3c$ space group with polarization along the [111]-direction (blue arrow). Due to the polar distortions, there are two inequivalent bondings between Nb-$d$ and O-$p$ orbitals, denoted with different colors.}
\end{figure}

Fig.~\ref{fig:atomic_str} presents the atomic structures of the ferroelectric BTO, PTO, and LNO. Each B-site ion is surrounded by a corner-sharing oxygen octahedron, with A-site ions filling the space between the oxygen octahedra. For both polar BTO and PTO, we consider polarization along [001] direction, resulting in a tetragonal $P4mm$ (No. 99) structure. Due to the polar distortion, the apical (O$_\parallel$) and equatorial (O$_\bot$) oxygen atoms occupy $1a$ and $2c$ Wyckoff positions, respectively. Therefore, there are three kinds of Ti-O bonds: two for the long (orange) and short (purple) bonds to apical oxygen atoms and the remaining one (gray) for the equatorial atoms. For LNO, we consider the rhombohedral R3c (No. 161) structure with polarization along the [111] direction, where there are two types of Nb-O bonds corresponding to the three short (purple) and three long (orange) bonds.

\begin{table}[htb!]
\caption{\label{tab:UVval} Calculated Hubbard parameters (in eV). $U$ and $U$+$V$ represent DFT+$U$ and DFT+$U$+$V$ calculations, respectively. $U_\text{d}$ is the on-site Hubbard parameter of the transition metal-$d$ orbitals. $U_\text{O}$ are the on-site Hubbard parameter of O-$p$ orbitals where the subscripts $\bot$ and  $\parallel$ are used for distinct O atomic sites (Fig.~\ref{fig:atomic_str} (a)). In BTO and PTO $V_{\text{d-O}_{\bot}}$, $V_{\text{d-O}_\parallel(S)}$, and $V_{\text{d-O}_\parallel(L)}$ are the inter-site Hubbard parameters between the Ti-$d$ and O-$p$ orbitals of O$_\bot$, O$_\parallel(S)$ and O$_\parallel(L)$, respectively. In LNO $V_{\text{d-O}(S)}$, and $V_{\text{d-O}(L)}$ are those between Nb-$d$ and O-$p$ orbitals of O$_\parallel(S)$ and  O$_\parallel(L)$, respectively (Fig.~\ref{fig:atomic_str} (b)).}
\begin{ruledtabular}
\begin{tabular}{ccccccccc}
      &\multicolumn{2}{c}{BaTiO$_3$}&\multicolumn{2}{c}{PbTiO$_3$}& &\multicolumn{2}{c}{LiNbO$_3$}\\ \cline{2-3} \cline{4-5}  \cline{7-8} 
      &$U$&$U+V$&$U$&$U+V$& & $U$ & $U+V$  \\      \hline
   $U_\text{d}$                   &0.24 &0.26 &0.17 &0.21 &$U_\text{d}$&0.29 &0.38 \\
$U_{\text{O}_\bot}$          &7.83 &7.73 &7.33 &7.06 &$U_\text{O}$&8.47 &8.28 \\
$U_{\text{O}_\parallel}$     &7.83 &7.36 &7.44 &7.12 &       &     &     \\
$V_{\text{d-O}_\bot}$        &     &2.36 &     &2.29 &       &     &     \\
$V_{\text{d-O}_\parallel(S)}$&     &2.53 &     &2.59 &$V_{\text{d-O}(S)}$&     &2.71 \\
$V_{\text{d-O}_\parallel(L)}$&     &2.04 &     &1.58 &$V_{\text{d-O}(L)}$&     &2.45

\end{tabular}
\end{ruledtabular}
\end{table}

Table~\ref{tab:UVval} presents the Hubbard parameters of the on-site and inter-site Coulomb interactions based on the ACBN0 functional and its extension~\cite{agapito2015,lee2020} denoted with $U$ and $U+V$, respectively. As the Hubbard parameters can vary depending on the atomic positions, the $U$ and $V$ values are re-calculated after the relaxation run, until the convergence is reached. There are small differences between the $U$ values of oxygen $p$ orbitals of BTO and PTO due to the polar distortions, except for the BTO calculated only with the on-site $U$ having the cubic structure without polar distortions, which will be discussed later in detail. In the case of LNO, all oxygen ions are symmetry-related and thus have the same on-site $U$ values. The calculated inter-site Hubbard parameters $V$ represent the interaction between the transition metal $d$ and oxygen $p$ orbitals, which are sensitive to the bond lengths having larger values for shorter bonds than longer bonds. This is due to the differences in the renormalized occupation number ${N}^{I\sigma}_{\psi_{m\mathbf{k}}}$ induced by the change in the occupation numbers in localized subspace surrounding each atom, increasing as the bond length decreases.

We note that the $U_{\text{d}}$ is less than 1 eV, which is significantly smaller than previously reported empirical values~\cite{Erhart2014,Tsunoda2019} or those derived from linear response theory (LRT)~\cite{Gebreyesus2023} and conversely, $U_{\text{O}}$ values are substantially larger than $U_{\text{d}}$. The $U_d$ and $U_{\text{O}}$ values are consistent with those obtained by ACBN0 method in ternary oxides~\cite{Jang2023prl} and TiO$_2$~\cite{agapito2015}. The differences in the interaction parameters can be understood by the way that ACBN0 functional treats the Hartree-Fock energy of frontier orbitals, giving more weight to highly occupied orbitals. In the case of BTO, the L\"{o}wdin charges of Ti and O are $0.45e$ and $1.72e$, respectively. It indicates that Ti-$d$ states contribute less to the Hartree-Fock treatment of the Coulomb energy than O-$p$ states. This is different from the LRT method~\cite{Timrov2021,Timrov2022}, which computes the Hubbard parameters through the changes in the occupation number via potential changes in the localized subspace. Despite the difference in $U$ values between ACBN0 and LRT schemes, the role of Hubbard correction, in essence, is suppressing charge fluctuations, either putting penalty for the fractional occupancy of O-$p$ orbitals (in our case) or Ti-$d$ orbitals (LRT), but not both as shown in Ref.~\citenum{Gebreyesus2023}. This ACBN0 scheme has been demonstrated to be effective in ternary oxides~\cite{Jang2023prl} and TiO$_2$~\cite{agapito2015}, where mainly putting a penalty on the O-$p$ orbitals can yield results closer to experimental values compared to standard DFT.

\begin{table}[htb!]
\setlength{\tabcolsep}{.1pt} 
\renewcommand{\arraystretch}{1.2} 
\caption{\label{tab:GeoChara} Structural parameters of relaxed atomic structures. V (\AA$^3$) and $dz_X$ (\AA) are the unit-cell volume and polar distortion of ion $X$ in Fig.~\ref{fig:atomic_str}, respectively. For BTO and PTO, $a$ (\AA) is the in-plane lattice constant, and $c/a$ is the ratio of the out-of-plane to the in-plane lattice constant. For LNO lattice constants and volume in the hexagonal setting are shown where  $a_H=2a_R\sin \frac{\alpha_R}{2}$ and $c_H=a_R \sqrt{3(1+2\cos\alpha_R)}$ with $a_R$ and $\alpha_R$ being the rhombohedral lattice constant and rhombohedral angle, respectively. $U$ and $U+V$ denote DFT+$U$ and DFT+$U$+$V$ methods, respectively, in which the Hubbard parameters are determined by the ACBN0 approach and its extension scheme.}
\begin{ruledtabular}
\begin{tabular}{cccccccc}
\multicolumn{7}{c}{BaTiO$_3$} \\
 & LDA    &PBEsol& $U$ & $U+V$ & HSE06 & Exp. \\ \hline
 $a$                      & 3.93  & 3.96  & 3.97  & 3.94  & 3.96  & 3.99 \footnotemark[1] \\
 $c/a$                    & 1.01  & 1.02  & 1.00  & 1.04  & 1.05  & 1.01 \footnotemark[1] \\
 V                        & 61.5  & 63.5  & 62.3  & 63.6  & 65.2  & 64.0 \footnotemark[1] \\
$dz_{\text{Ti}}$           & 0.049 & 0.064 & 0 & 0.067 & 0.082 & 0.060\footnotemark[1] \\
$dz_{\text{O}_\parallel}$ &-0.083 &-0.123 & 0 &-0.178 &-0.199 &-0.093\footnotemark[1] \\
$dz_{\text{O}_\bot}$      &-0.054 &-0.076 & 0 &-0.111 &-0.105 &-0.056\footnotemark[1] \\\hline\hline
\multicolumn{7}{c}{PbTiO$_3$} \\
 & LDA    &PBEsol& $U$ & $U+V$ & HSE06 & Exp. \\ \hline
 $a$                      & 3.86  & 3.86  & 3.90  & 3.80  & 3.81  & 3.90 \footnotemark[2] \\ 
 $c/a$                    & 1.05  & 1.10  & 1.02  & 1.23  & 1.22  & 1.07 \footnotemark[2] \\ 
 V                        & 60.1  & 63.3  & 61.0  & 67.3  & 67.6  & 63.3 \footnotemark[2] \\ 
$dz_{\text{Ti}}$           & 0.139 & 0.171 & 0.143 & 0.275 & 0.261 & 0.157\footnotemark[2] \\ 
$dz_{\text{O}_\parallel}$ & 0.378 & 0.533 & 0.306 & 0.878 & 0.864 & 0.465\footnotemark[2] \\ 
$dz_{\text{O}_\bot}$      & 0.426 & 0.545 & 0.352 & 0.789 & 0.771 & 0.488\footnotemark[2] \\\hline\hline
\multicolumn{7}{c}{LiNbO$_3$} \\
 & LDA    &PBEsol& $U$ & $U+V$ & HSE06 & Exp. \\ \hline
 $a_H$                      & 5.07  & 5.13   & 5.10  & 5.11  & 5.13 & 5.15 \footnotemark[3] \\
 $c_H/a_H$                    & 2.70  & 2.69   & 2.69  & 2.70  & 2.70 & 2.69 \footnotemark[3] \\
 V                        & 304.2 & 315.1  & 309.1 & 312.6 & 315.2& 318.8\footnotemark[3] \\
$dz_{\text{Nb}}$           & 0.460 & 0.447 & 0.454 & 0.440 & 0.440 & 0.457\footnotemark[3] \\
$dz_{\text{O}}$           & 0.705 & 0.709 & 0.641 & 0.715 & 0.715 & 0.723\footnotemark[3] 
\end{tabular} 
\end{ruledtabular}
\footnotetext[1]{Room temperature data from Ref.~\cite{Shirane1957}}
\footnotetext[2]{Room temperature data from Ref.~\cite{Shirane1956}}
\footnotetext[3]{300 K data form Ref.~\cite{Boysen1994}}
\end{table}

\begin{figure*}[htbp!]
\includegraphics[width=0.8\textwidth, angle=-0]{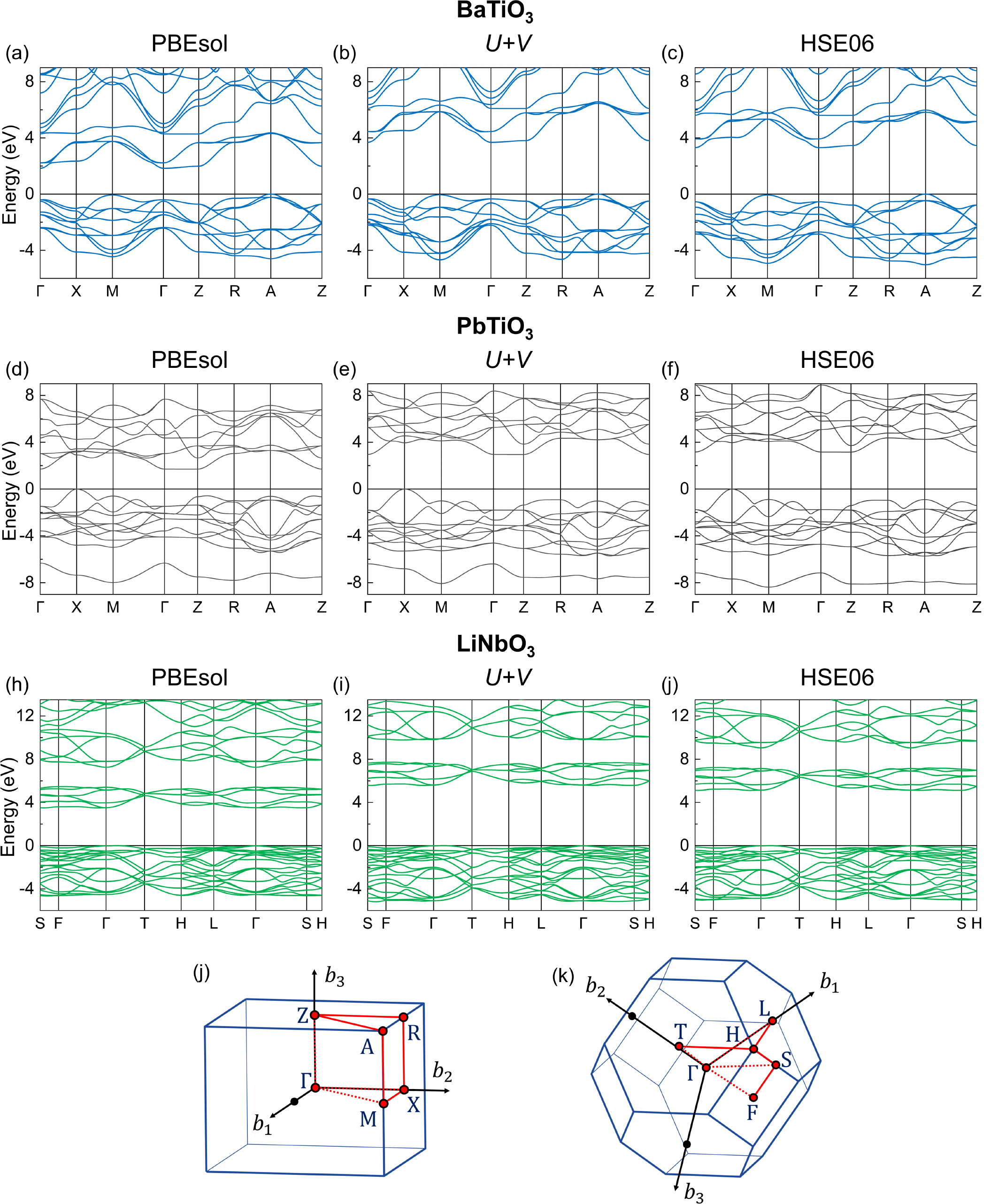}
\caption{\label{fig:bands} Band structures of ferroelectric phases along high-symmetry lines.  Band structures of (a-c) tetragonal BTO, (d-f) tetragonal PTO, (g-i) rhombohedral LNO. The panels (a,d,g), (b,e,h), and (c,f,i) are calculated with PBEsol, DFT+$U$+$V$ ($U+V$), and HSE06 schemes, respectively. The Brillouin zone of (j) tetragonal and (k) rhombohedral unit cell.}
\end{figure*}

We investigate how the different treatments of Coulomb interaction have an effect on polar distortions. Table~\ref{tab:GeoChara} presents the structural parameters calculated with LDA, PBEsol, DFT+$U$, DFT+$U$+$V$, and HSE06. We find that the inclusion of only the on-site Hubbard interaction suppresses the polar distortions, having significantly smaller values compared with those from other functionals and experimental data. Adding the inter-site Hubbard term increases the polar distortions substantially. We note that the resulting polar distortions are close to those from HSE06, suggesting that including on-site and inter-site Hubbard interactions gives similar corrections to HSE06.

It is worth discussing the absence of polar distortions of BTO in our DFT+$U$ calculation. This is not due to the large $U$ value on O-$p$ orbital as it is also shown that the DFT+$U$ calculations with LRT parameterization also do not exhibit polar instability \cite{Gebreyesus2023}. In both cases, the on-site Coulomb interaction suppresses the covalency too strongly to induce the polar instability, and the instability can be restored by the inclusion of the intersite Coulomb interaction that enhances covalency \cite{Gebreyesus2023}. Again, either putting the $U$ value mainly on O-$p$ or Ti-$d$ does not make a qualitative difference, but the inclusion of the $V$ is key to restoring the polar instability. In contrast the persistent polar distortion of PTO and LNO in the $U$-only scheme is due to the significant A-site polar distortions, contributing to the polar instability. Unlike BTO, having the polar distortions dominantly driven by shifting Ti ions \cite{Zhong1995}, PTO has comparable Pb and Ti distortions \cite{Waghmare1997}, and LNO has polar distortion mainly by the shift of Li atom \cite{Parlinski2000}. Thus, the effect of $U$ substantial enough to suppress the polar instability of BTO is weaker in PTO and even smaller in LNO as the contribution of the A-site motion becomes dominant than the B-site. 

In all the compounds considered, the DFT+$U$+$V$ calculations tend to increase the degree of polar distortions compared with LDA and PBEsol results, and interestingly, the resulting polar distortions are closest to HSE06 results. The former suggests that the increased covalency from the inter-site $V$ terms compensates the charge localization effect of on-site $U$ correction and that the inter-site interaction should be included to reproduce the correct polar ground states. The latter indicates that including the Coulomb interaction up to the nearest-neighbor inter-site interaction yields the polar distortions comparable to HSE06 results despite the different methodologies in treating the long-range Coulomb interaction between DFT+$U$+$V$ and HSE06 schemes. We note that for LNO, the polar distortions computed with DFT+$U$+$V$ show almost negligible differences compared with PBEsol, but only when both $U$ and $V$ are included simultaneously, as calculations with $U$ alone result in substantially decreased oxygen distortions; this again stresses the importance of including the inter-site interaction. Overall, all exchange-correlation functionals, except for the $U$-only scheme, produce reasonable polar distortions that are comparable with experimental data, with both $U$+$V$ and HSE06 schemes showing slightly increased polar distortions.

\subsection{\label{subsec:band_gaps} Electronic structures}

The band structures of each material in the ferroelectric phase, presented in  Fig.~\ref{fig:bands}, clearly show the most prominent effect from the extended Hubbard interaction, which is substantially increased band gap values compared with PBEsol calculations. (See SI Fig. S1 for band structures with the other exchange correlational functionals.) We observe an increase in the direct band gap by approximately 
1.5$\sim$2.0 eV, while changes in bandwidth remain relatively small. This behavior can be attributed to enhanced polar distortions in both the DFT+$U$+$V$ and HSE06 calculations, which reduce the $p$-$d$ hopping character. This reduction offsets the bandwidth increase caused by inter-site interactions. Overall, the band dispersions obtained with DFT+$U$+$V$ are similar to those from HSE06, consistent with their similar structural parameters. 

\begin{figure}[htbp!]
\includegraphics[width=0.8\columnwidth, angle=-0]{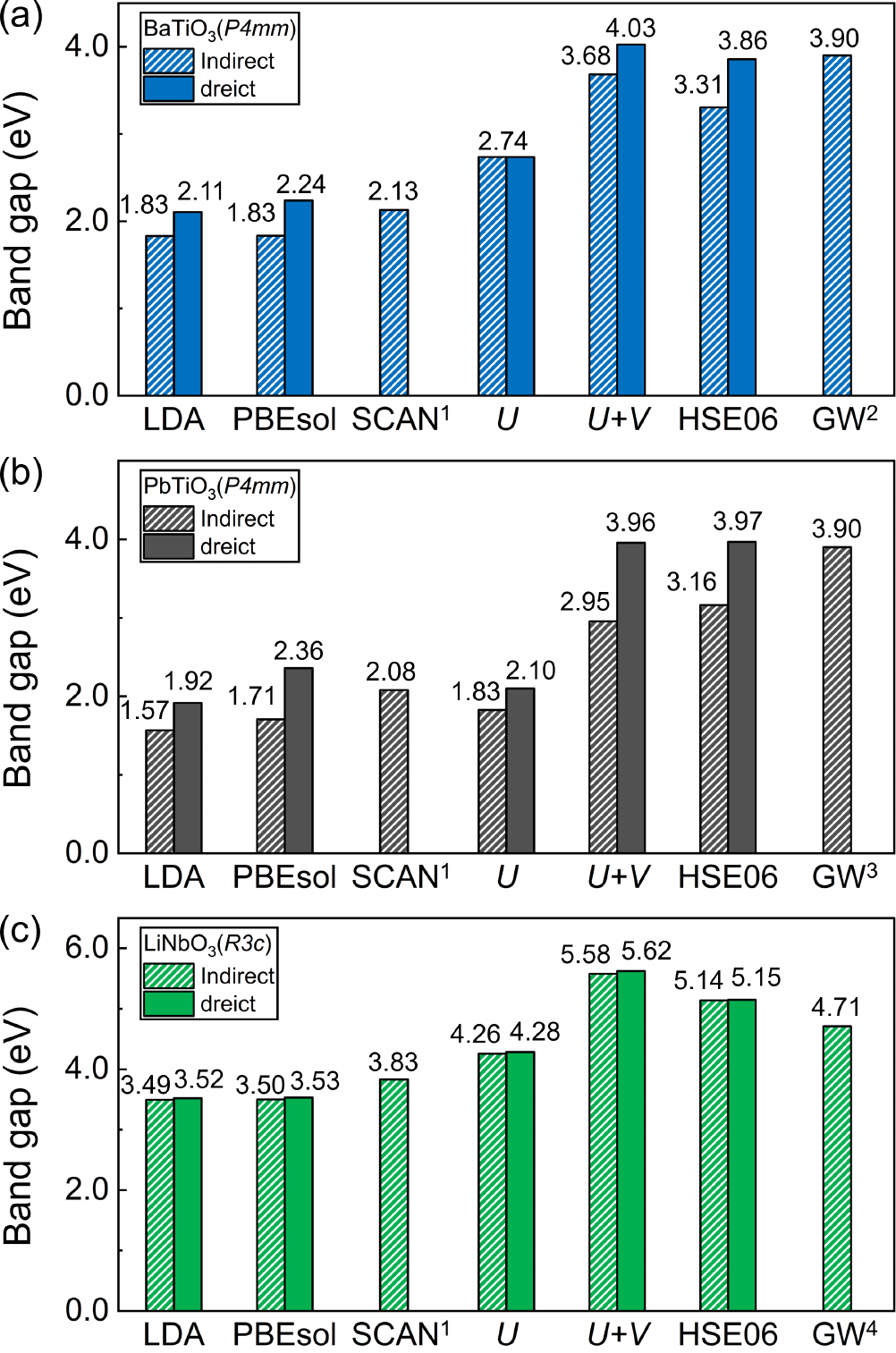}
\caption{\label{fig:epsart} Band gap values depending on the exchange-correlation functional. The band gap values of (a) BTO and (b) PTO and (c) LNO calculated with LDA, PBEsol, SCAN, DFT+$U$ ($U$), DFT+$U$+$V$ ($U+V$), HSE06, and $GW$ are presented. The solid and hatched bars denote direct and indirect band gaps, respectively. The  SCAN$^1$ and $GW^{2-4}$ data are from Ref.~\citenum{Zhang2017} and  Refs.~\citenum{Sanna2011,Gou2011,Thierfelder2010}, respectively.}
\end{figure}

Fig.~\ref{fig:epsart} summarizes the calculated indirect and direct band gap values in the ferroelectric phases, along with values obtained with SCAN and $GW$  from the literature ~\cite{Zhang2017,Sanna2011,Gou2011,Thierfelder2010}. We find the DFT+$U+V$ calculations substantially increase the band gap values in which most of the band gap enhancements are from inter-site $V$ term. As the results, the band gap values calculated with DFT+$U$+$V$ are in good agreement with ideal $GW$ values \cite{Sanna2011, Gou2011, Thierfelder2010} and also similar to HSE06 values. We note that the band gap values using both LDA and PBEsol functionals substantially underestimated values~\cite{Zhang2017} with SCAN yielding a small band gap enhancement not sufficient to match the $GW$ ones. The experimental band gaps of BTO, PTO, and LNO are 3.27, 3.60, and 3.45 eV~\cite{Wemple1970,Zametin1984,Dhar1990}, respectively. The band gaps obtained from DFT+$U$+$V$, HSE06, and $GW$ calculations overestimate these experimental values by approximately 0.3 to 1.0 eV because these computational methods do not fully account for excitonic effects present in the materials~\cite{Eglitis2002BTO,Eglitis2002PTO,Schmidt2021}.

\subsection{\label{subsec:ferro_pro} Ferroelectric properties}
\begin{figure}[htbp!]
\includegraphics[width=0.8\columnwidth, angle=-0]{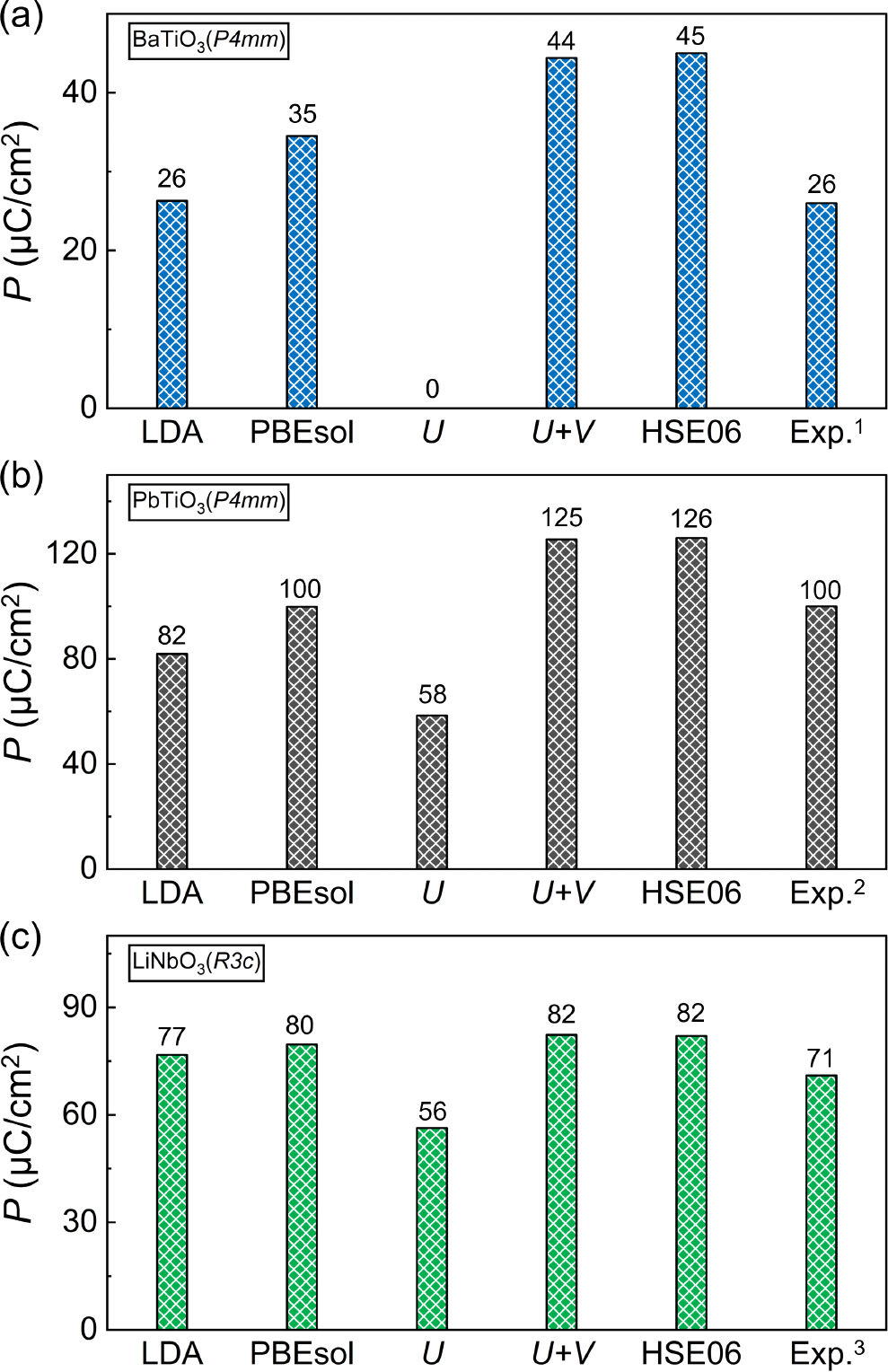}
\caption{\label{fig:polar} Calculated polarization values depending on choice of exchange-correlation functionals. The polarization values of (a) BTO, (b) PTO, and (c) LNO calculated with LDA, PBEsol, DFT+$U$ ($U$), DFT+$U$+$V$ ($U+V$), and HSE06 are compared with experimental data$^{1-3}$ from Refs.~\citenum{Wieder1955,Lines2001,Glass1976}, respectively.}
\end{figure}

We investigate the dependence of ferroelectric properties on the choice of exchange-correlation functionals, in which we focus on polarization, switching energy barrier, and Born effective charges (BECs). Fig.~\ref{fig:polar} presents the calculated polarizations compared with experimental data \cite{Wieder1955,Lines2001,Glass1976}. We find that except for the DFT+$U$ calculations showing substantially suppressed polarization or incorrect zero polarization, all methods produce values in good agreement with experimental data. We find that the polarization values calculated with DFT+$U$+$V$ and HSE06 show slight overestimation compared with the experimental data, which is due to the relatively larger polar distortions calculated with DFT+$U$+$V$ and HSE06. Again, similar polarization values are calculated with DFT+$U$+$V$ and HSE06, consistent with similar structural parameters and band structures.

Table~\ref{tab:ElcChara} presents the component of BEC tensor along the polarization direction, revealing interesting exchange-correlation functional dependencies. First of all, the results obtained with DFT+$U$ clearly show the reduced magnitude of the BEC values due to the decreased covalency, which are compensated by the inclusion of the inter-site $V$ interaction, resulting in similar BEC values between PBEsol and the DFT+$U$+$V$ method. Interestingly, the BEC values obtained with DFT+$U$+$V$ are slightly larger than those from HSE06, except for slightly smaller A-site values. The slight decrease in the BEC of the A-site is due to the fact that DFT+$U$+$V$ calculations do not account for A-site orbitals. This suggests that additional corrections for A-site orbitals are unnecessary for the materials studied. In addition, we observe a clear trend in $E_b$, defined as the energy difference between the ferroelectric and centrosymmetric structures. The value of $E_b$ increases in the following order: LDA, PBEsol, HSE06, and DFT+$U$+$V$. We note that the DFT+$U$ result is omitted due to its centrosymmetric ground state. The increase in LDA to PBEsol is related to the larger lattice constants that enhance the polar instability, while the greater polar distortion of the HSE06 scheme is the main cause of its enhancement $E_b$. The extended Hubbard interaction further increases $E_b$ over the HSE06 results by strengthening the adjacent bonding through the enhanced covalency, which is consistent with the relatively higher BEC values.

\begin{table}[htbp!]
\setlength{\tabcolsep}{6pt} 
\renewcommand{\arraystretch}{1.1} 
\caption{\label{tab:ElcChara} Born effective charges $Z^*$ calculated at $Pm\bar{3}m$ (BTO and PTO) and $R\bar{3}c$ (LNO) phases, spontaneous polarization $P$ ($\mu C/cm^{2}$) of the fully relaxed ferroelectric phases, the total energy difference $E_b$ (meV) between the optimized centrosymmetric and ferroelectric phases of BTO, PTO and LNO.}
\begin{ruledtabular}
\begin{tabular}{cccccc}
\multicolumn{6}{c}{BaTiO$_3$}  \\
 & LDA    &PBEsol& $U$ & $U+V$ & HSE06  \\ \hline
$Z^*_\text{A}$             & 2.76  & 2.76  & 2.69  & 2.65  & 2.68 \\
$Z^*_\text{B}$             & 7.27  & 7.27  & 6.33  & 7.12  & 6.40 \\
$Z^*_{\text{O}_\parallel}$ & -5.72 & -5.75 & -4.82 & -5.48 &-5.14 \\
$Z^*_{\text{O}_\bot}$      & -2.16 & -2.14 & -2.10 & -2.14 &-1.97 \\
$E_b$                      & 1.2 & 3.5 & - & 30.2 & 11.1 \\\hline\hline
\multicolumn{6}{c}{PbTiO$_3$}  \\
 & LDA    &PBEsol& $U$ & $U+V$ & HSE06  \\ \hline
$Z^*_\text{A}$             & 3.91  & 3.89   & 3.48  & 3.44  & 3.79 \\
$Z^*_\text{B}$             & 7.13  & 7.13   & 6.33  & 7.16  & 6.44 \\
$Z^*_{\text{O}_\parallel}$ & -5.83 & -5.86  & -4.96 & -5.63 &-5.24 \\
$Z^*_{\text{O}_\bot}$      & -2.60 & -2.58  & -2.42 & -2.48 &-2.50 \\
$E_b$                      & 12.0 & 17.2 & 36.2 & 126.6 & 39.6\\ \hline\hline
\multicolumn{6}{c}{LiNbO$_3$} \\
 & LDA    &PBEsol& $U$ & $U+V$ & HSE06  \\ \hline
$Z^*_\text{A}$             & 1.10  & 1.10  & 1.07  & 1.06  & 1.11  \\
$Z^*_\text{B}$             & 9.07  & 9.16  & 7.88  & 8.84  & 8.31  \\
$Z^*_{\text{O}}$           & -3.39 & -3.42 & -2.98 & -3.30 & -3.12 \\
$E_b$                      &25.8 & 26.9  & 29.9 & 132.2 & 61.4\\
\end{tabular}
\end{ruledtabular}

\end{table}

\section{Summary}
We investigate the electronic and ferroelectric properties of three representative ferroelectric materials, BTO, PTO, and LNO, incorporating both the on-site and inter-site Hubbard interactions based on the extended ACBN0 method, which enables a self-consistent determination of Hubbard parameters. Various material properties such as band structures, ferroelectric distortions, polarization, BECs, and switching barriers are calculated and compared with other exchange-correlation functionals, including local (LDA), semi-local (PBEsol), hybrid (HSE06), and on-site Hubbard functionals. For the evaluated properties, most of the exchange-correlation functionals predict values reasonably well compared with experimental data, whereas we find the DFT+$U$ calculation predicts the incorrect centrosymmetric ground-state of BTO and substantially reduced polar distortions and BECs for other materials. This is due to the excessive reduction of the covalency by considering the on-site Hubbard interaction alone, which can be remedied by including inter-site Hubbard terms, improving most of the predicted values. Among the properties calculated using the DFT+$U$+$V$ method, the band gap values stand out for their superior accuracy compared to those obtained with local and semilocal exchange-correlation functionals. The inclusion of inter-site Hubbard corrections is essential, as it enables band gap predictions that closely align with $GW$ calculation results. This is particularly important for predicting the efficiency of shift current photovoltaic effect and electronic reconstruction at ferroelectric heterojunctions, requiring accurate prediction of the band gap values and band alignments, respectively. 

Overall, we find that the calculated properties with the DFT+$U$+$V$ method are surprisingly close to those from HSE06, which is interesting in two aspects. One is the similarity of results despite the different treatment of the Hartree-Fock type correction, one taking Hubbard type of Coulomb interaction up to a certain distance with double-counting correction and the other taking the mixture of semilocal and screened exchange terms. We believe that self-consistency in determining the Hubbard parameters plays a crucial role in giving the results insensitive to the detail of the functional form, as long as a sufficient range of Hubbard interactions is included; this is the subject of our future study. The other is the drastically reduced time for the DFT+$U$+$V$ method compared with that using HSE06. From our calculation environment (see SI section 2 for details), the time required for reaching the electronic convergence for the ground-state structure is about 30 times smaller for the DFT+$U$+$V$ method compared with HSE06. Thus, we find that the use of self-consistently determined on-site and inter-site Hubbard parameters is greatly beneficial for high-throughput calculations or systems with large numbers of atoms where we expect the material properties similar to HSE06 results calculated with much reduced calculation cost.

\subsection*{Author Contributions}
S.Y.P. and Y.-W.S. proposed and designed research. M.C. and W.Y. performed first-principles calculations. M.C., W.Y., Y.-W.S., and S.Y.P. wrote the manuscript. All authors read and approved the final manuscript.

\subsection*{Acknowledgments}
S.Y.P. was supported by the National Research Foundation of Korea(NRF) grant funded by the Korea government(MSIT) (No. RS-2024-00358551) and by the Basic Science Research Program through the National Research Foundation of Korea(NRF) funded by the Ministry of Education (No. 2021R1A6A1A03043957). S.Y.P and M.C.C. were supported by the Basic Science Research Program through the National Research Foundation of Korea(NRF) funded by the Ministry of Education (No. 2021R1A6A1A10044154).
Y.-W.S. was supported by KIAS individual Grant (No. CG031509). W.Y. was supported by KIAS individual Grant (No. QP090102).
Computations were also supported by the CAC of KIAS.

\subsection*{Competing interests}
The authors declare no competing interests.

\subsection*{Data availability}
The datasets generated and/or analysed during the current study are available from the corresponding author on reasonable request.

\nocite{*}


\setcounter{table}{0}
\renewcommand{\thetable}{S\arabic{table}}%
\setcounter{figure}{0}
\renewcommand{\thefigure}{S\arabic{figure}}%
\setcounter{equation}{0}
\renewcommand{\theequation}{S\arabic{equation}}%

\renewcommand*{\thesubsection}{\arabic{subsection}}\setcounter{subsection}{0}

\hfill\break

\section*{Supplementary Materials}

\subsection{The band structures calculated with the LDA and PBEsol+$U$ functionals}

\begin{figure}[H]
\includegraphics[width=1.0\columnwidth, angle=-0]{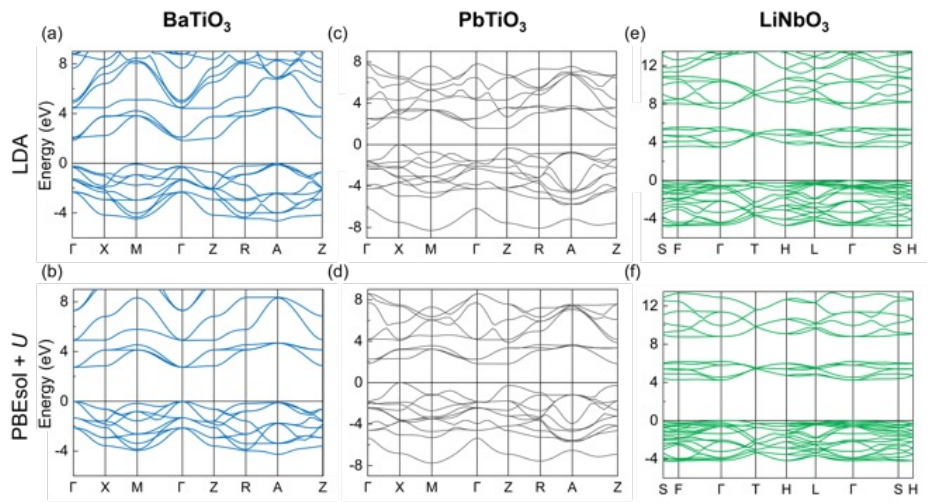}
\caption{\label{fig:supp1} 
Comparison of the band structures calculated with the LDA (panels (a, c, e)) and PBEsol+$U$ (panels (b, d, e)) functionals. The band dispersions of the relaxed structures of the (a) tetragonal and (b) cubic BTO, (c-d) tetragonal PTO, and (e-f) rhombohedral LNO are presented. The high symmetry lines and points in the Brillouin zone are presented in Fig. 2.
}
\end{figure}

\subsection{The comparison of computation time between DFT+$U$+$V$ and HSE06 methods}

We compare the efficiency of the DFT+$U$+$V$ based on the extended ACBN0 method against the HSE06 hybrid functionals by measuring their computation times. We perform self-consistent calculations on the ferroelectric structure of BTO using 32 CPU cores with the calculational parameters stated in Sec. III (computational details). The time required to reach the energy difference of consecutive self-consistent loops below 10-8 Ry is 4 minutes for the DFT+$U$+$V$ method, which shows a 31-fold reduction compared with 125 minutes for the HSE06 functional. Similarly, the average time per self-consistent iteration (2.2 seconds) is reduced by a factor of 214 compared with the HSE06 functional (471 seconds).

\end{document}